\documentclass[12pt,a4paper]{article}
\usepackage{latexsym}

\title{
\large\bf Quantum mechanics in the $K$-Field formalism:\\
  the basic principles of geometrization}
\author{\large K.B. Korotchenko}
\date{}
\begin{document}
 \maketitle
\begin{abstract}
 The basic principles of the quantum mechanics in the K-field formalism
 are stated in the paper. The basic distinction of
 this theory arises from that the quantum theory equations
 (including well-known Schr\"{o}dinger, Klein-Gordon and
 quadratic Dirac equations) are obtained
 from de Broglie postulate geometric generalization.
 Rather, they are obtained as the free wave equations on a manifold
 metrizing force interactions of particles.\\
 Such view on the quantum theory basic equations allows one
 to use semiclassical models for the quantum system simulation.
 The quantization principle modifies as well. Namely,
 quantum system stationary conditions are such conditions,
 at which test particles motion is Lyapunov stable.
\end{abstract}

 n1. The basic ideas of the quantum theory geometric formalization
 (i.e. the $K$-field formalism) are presented in the papers \cite{p1}.
 However, in the first papers devoted to the problem of a geometric
 correlation between the classical and the quantum mechanics,
 the author did not plan to construct the quantum theory formalism.
 Therefore, in the referred papers \cite{p1} it was not always
 possible to find the direct answers to problems originating
 at acquaintance with the $K$-field formalism.

 By virtue of this (and answering on requests of some correspondents)
 the author considers that the more in-depth account of the quantum
 mechanics basic positions in the K-field formalism is expedient.

 n2. The modern classical statement of the de Broglie postulate
 establishes possibility of the correlation between a particle
 and some wave process.
 Such a conception of the postulate is very convenient for
 the quantum theory classical formalism, because it allows to build
 this formalism completely irrespective of the de Broglie postulate.

 However, the de Broglie hypothesis initial sense is that
 it is possible to spread the formulas describing a photon
 behaviour to mass particles too.
 Such a view on the de Broglie theory allows one to use it for a
 formal description of a microparticles behaviour in terms of a
 photon behaviour formal description.

 Such a conception of the de Broglie hypothesis we shall accept as
 a basis of the quantum theory formalism.

 n3. To present correlation of photon and microparticle description
 more obviously let's make a table (see tab. 1).
\begin{center}
\begin{tabular}{||p{62 true mm}|p{62 true mm}||}
\multicolumn{2}{r}{Table 1}\\
\hline
\multicolumn{1}{||c|}{photons} &
\multicolumn{1}{c||}{mass particles} \\
\hline
\hline
 move uniform rectilinearly with the velocity $c $ relatively
 to any inertial frame &
 move along any trajectories with any velocities (less than $c $)
 relatively to any inertial frame\\
\hline
 the wave properties are described by the equation of d'Alember
\[ \Box A^{\mu}\: =\: 0\: \]
 for an electromagnetic field four-potential $A^\mu $ &
 the wave properties are described by the Klein-Gordon equation
\[ [(E - V)^2 + \hbar^2 c^2 \triangle - m^2 c^2] \Psi\: =\: 0 \]
 for a particle $\Psi$-function \\
\hline
\hline
\end{tabular}
\end{center}
\ \\
 Obviously by virtue of such essential difference in description of
 photons and microparticles the direct prolongation of
 the equations for photons on mass particles is impossible.

 Let's look at this problem a little differently.
 Let's consider motion of particles from an isolated
 observer's view point (it will allow to avoid problems connected
 with transition from one reference frame to another).

 Photons move along an isotropic geodesic lines of the Minkowski
 four-space $V_4 $.
 Hence, it is necessary to build a four-space in which mass particles
 should move (relatively to an isolated observer) along
 an isotropic geodesic lines of this space.

 Let's now collect all the isotropic surfaces of separate observers
 in one four-manifold $^kV_4 $.
 We shall obtain the manifold $^kV_4 $ in which
 mass particles will move (relatively to any observer) along
 isotropic geodesic lines of this manifold for any motion.

 Hence, description of a mass particle on the manifold
 $^kV_4 $ and description of a photon in the Minkowski
 four-space $V_4 $ becomes equivalent.
 And so, we can formulate the de Broglie hypothesis as
 follows:
\begin{itemize}
\item[\ ] to describe wave properties of particles it is necessary
\begin{itemize}
\item[\ -] to build the manifold $^kV_4 $ in which mass particles
 move along geodesic lines of this manifold for any motions;
\item[\ -] to build an operator the similar to that d'Alember on the
 manifold $^kV_4 $
 (so-called the de Rham operator $^{(k)}\triangle $);
\end{itemize}
\item[\ ]
 then the equation
\begin{equation}
\label{e01}
 ( ^{(k)}\triangle k )_{\mu}\: =\: 0\: ,
\end{equation}
 where $k_{\mu}$ is the $K$-field potential should describe
 the wave properties of mass particles.
\end{itemize}
 Let's sum up everything, mentioned above, in a table (see tab. 2)
\begin{center}
\begin{tabular}{||p{62 true mm}|p{62 true mm}||}
\multicolumn{2}{r}{Table 2}\\
\hline
\multicolumn{1}{||c|}{photons} &
\multicolumn{1}{c||}{mass particles} \\
\hline
\hline
 move along isotropic geodesic lines of the Minkowski
 four-space $V_4 $ &
 move along isotropic geodesic lines
 of the manifold $^kV_4 $\\
\hline
 the wave properties are described by the equation of d'Alember
\[ \Box A^{\mu}\: =\: 0\: \]
 for the electromagnetic field four-potential $A^\mu $ &
 the wave properties are described by the equation
\[( ^{(k)}\triangle k )_{\mu}\: =\: 0\: \]
 for the $K$-field potential $k_\mu $, where $^{(k)}\triangle $
 is the de Rahm operator and $k_{\mu} $ is linear form \\
\hline
\hline
\end{tabular}
\end{center}

 n4. To describe wave properties of particles to content ourself
 in this paper with construction special geometric formulation of
 force interactions.

     Geometrization of an
     interaction consists in finding a metric space in
     which the test particle trajectories are
     geodesic lines \cite{p2}. This is the starting point of
     Einstein concept of geometrization.

     An interesting method of metrization of arbitrary
     force interactions corresponding to this concept
     was presented in \cite{p3}. In this
     method of metrization, the test particles move
     along geodesic lines.  However, the force fields are related
     with the components of the connection tortion tensor
     of a pseudo-Euclidean space.  In this sense, the
     metrization of force interactions presented in \cite{p3} does
     not correspond to the Einstein concept because the
     metric properties of the space do not depend on force fields.

     So, we shall consider a metric
     statement of force interactions in which, as in \cite{p3},
     the test particle motion equations
     represent a special form of Newton's second law in
     four-dimensional form but the metric tensor and physical
     fields are interdependent.

     To avoid the problems connected with
     the distinction between the concepts of a reference
     frame and a coordinate system \cite{p3}, different observers
     (i.e., reference frames) will be associated with different
     isotropic surfaces on the manifold $^kV_4 $.

 n5. The states of the test particles (of mass $m $
     and charge $e $) in potential fields will be called
     classical states.  Correspondingly, all the
     characteristics of the particle describing its
     behavior in the classical state (trajectory, velocity,
     momentum, energy, etc.) will be called classical.

     It should be emphasized that all classical
     characteristics should be measured relatively to one
     specific reference frame. Any inertial frame (IF)
     may be selected as that reference frame.

     Let's consider some a four-dimensional space with
     the metric
\begin{equation}
\label{e1}
 ^{(k)}dS^2\: =\: {^{(k)}}g_{oo}(x^i, t) c^2 dt^2\: +
               \: g_{ik} dx^i dx^k\: ,
\end{equation}
     where ($- g_{ik} $) is the metric tensor of the Euclidean space
     $V_3 $.

     Any classical trajectory
     $x^i = x^i (t) $ may be considered as a line defined by the
     equation ${^{(k)}}g_{oo}(x^i (t), t)c^2 dt^2 =
      -þ g_{ik} dx^i dx^k $. And so along the line
\begin{equation}
\label{e2}
 {^{(k)}}g_{oo}(x^i (t), t)\: =\: v^i v_i/c^2\: ,
\end{equation}
     where $v^i $ is the particle velocity measured relatively to the
     specified IF ($v_i = þ- g_{ik} v^k $).

     Thus, each point $p $ of the classical particle
     trajectory $x^i = x^i (t) $ in $V_3 $ may be considered
     as a line lying on the isotropic surface
     $^kG_{o3} \subset {^k}V_4 $ described by the equation
     ${^{(k)}}g_{oo}c^2 dt^2 = -þ g_{ik} dx^i dx^k $.

     Hence, each point $p \in V_3 $ may also be
     considered as a point of the isotropic surface
     $^kG_{o3} $ in $^kV_4 $.
     That is an isotropic surface
     $^kG_{o3} \subset {^k}V_4 $ may be
     constructed at points of space $V_3 $. By changing
     the values of the initial parameters, a set of points
     covering the whole of $^kG_{o3} $ may be obtained.
     And by transiting from one reference frame to another, a set
     of surfaces $^kG_{o3} $ covering the whole of
     $^kV_4 $ may be obtained.
     That is an imbedding \cite{p2} (enclosure in a space of higher
     dimensionality) may be constructed.

 n6. According to Eq.(\ref{e1}), the method of enclosure
     described in Sec.5 should have the distinctive
     property. Namely, the geometry of the enclosing space
     $^kV_4 $
     should have no influence on the geometric properties
     of the enclosed space $V_3 $ (should not change
     the metric tensor $g_{ik} $). In other words, the imbedding
     must occur at those points of $^kV_4 $ at which the
     external curvature of the enclosed surface is zero.

     So, then it follows from the Gauss-þVaingarten equations
     (see \cite{p4}, for example), the
     absolute differential of the space $^kV_4 $ (denoted by
     ${^{(k)}}\nabla (\cdots) $) is defined by the equation
\begin{equation}
\label{e3}
 ^{(k)}\nabla A^{\mu}\: =\: ({^{(3)}}\nabla_i A^{\mu}) dx^i\: +
                       \: ({^{(4)}}\nabla_o A^{\mu}) dx^o\: .
\end{equation}
     Equation (\ref{e3}) may also be rewritten in the form
\begin{equation}
\label{e4}
 ^{(k)}\nabla A^{\mu}\: =\: {^{(k)}}D A^{\mu}\: +
     \: {^{(k)}}\Gamma^{\mu}_{\nu o} A^{\nu} dx^o\: ,
\end{equation}
     where ${^{(k)}}DA^i = DA^i + {^{(k)}}S^i_{k l}A^k dx^l $
     is the absolute differential of the Euclidean space
     $V_3 $ (here ${^{(k)}}S^i_{k l} = S_{k l}{^i}
     - S_l{^i}{_k} - S_k{^i}{_l} $
     and $S_{k l}{^i} $ is the tortion tensor) and
     ${^{(k)}}\Gamma^{\mu}_{\nu o} $ is the connection of the space
     $^kV_4 $.

 n7. To obtain a more detailed description, the
     definition of the absolute differential
     $^{(k)}\nabla (\cdots) $ is written in standard form
     \cite{p2}, \cite{p5}
\begin{equation}
\label{e5}
 ^{(k)}\nabla A^{\mu}\: =\: (\partial_{\nu} A^{\mu}\: +
     \: \Gamma^{\mu}_{\omega \nu} A^{\omega}) dx^{\nu}\: ,
\end{equation}
     where $2\Gamma^{\mu}_{\omega \nu} =
     2 {^{(k)}}\Gamma^{\mu}_{\omega \nu} +
     Q^{\mu}_{\omega \nu} $,
     at that $Q^{\mu}_{\omega \nu} =
     {^{(k)}}g^{\mu \gamma} ({^{(k)}}Q_{\omega\nu\gamma} +
     {^{(k)}}Q_{\nu\gamma\omega} -
     {^{(k)}}Q_{\gamma\omega\nu}) $ and
     ${^{(k)}}Q_{\mu\nu\omega} =
     - {^{(k)}}\nabla_{\mu} \Big({^{(k)}}g_{\nu \omega} \Big) $;
     ${^{(k)}}\Gamma^{\mu}_{\omega \nu} =
     \Big\{^\mu_{\omega\nu} \Big\} +
     {^{(k)}}S^{\mu}_{\omega\nu} $,
     where $\Big\{^\mu_{\omega\nu} \Big\} $ is the Christoffel symbol
     and ${^{(k)}}S^{\mu}_{\omega\nu} = S_{\omega\nu}{^{\mu}}
     - S_{\nu}{^{\mu}}{_{\omega}} - S_{\omega}{^{\mu}}{_{\nu}} $
     at that $S_{\omega\nu}{^{\mu}} $ is the tortion tensor.

     If it is required that the definition in Eq.(\ref{e5})
     coincide with that in Eq.(\ref{e4}), the result obtained is
\begin{equation}
\label{e6}
 2{^{(k)}}\Gamma^i_{o j} dx^j\: +\:
    Q^i_{o \omega}dx^{\omega}\: =\:
    Q^i_{j \omega}dx^{\omega}\: =\:
 2{^{(k)}}\Gamma^o_{\mu j} dx^j\: +\:
    Q^o_{\mu \omega} dx^{\omega}\: =\: 0\: ,
\end{equation}
     which must be satisfied if the imbedding described in
     Secs.5 and 6 is possible.

     It may readily be demonstrated that the absolute
     differential $^{(k)}\nabla (\cdots) $ of space $^kV_4 $
     defined by Eq.(\ref{e4}) describes a nonmetric transfer
     in $^kV_4 $. In fact
\begin{equation}
\label{e7}
    {^{(k)}}Q_{o o o}\: =\:
   2{^{(k)}}g_{o o}{^{(k)}}S^o_{o o}\: ,\; \;
    {^{(k)}}Q_{i o o}\: =\: - \partial_i {^{(k)}}g_{o o}\: .
\end{equation}
     The remaining ${^{(k)}}Q_{\mu\nu\omega} = 0 $. As a result
     the equations (\ref{e6}) take the form
\begin{eqnarray}
\label{e8}
 {^{(k)}}S^o_{o j} dx^j & = & - \Big\{^o_{o j}\Big\} dx^j -
 2{^{(k)}}S^o_{o o} dx^o\: ,\nonumber \\
\label{e8-1}
 {^{(k)}}S^o_{i j} dx^j & = & -  \Big\{^o_{i j}\Big\} dx^j +
 \Big\{^o_{i o}\Big\} dx^o\: ,\\
\label{e8-2}
 {^{(k)}}S^i_{o j} dx^j & = & -  \Big\{^i_{o j}\Big\} dx^j +
 \Big\{^i_{o o}\Big\} dx^o\: ,\nonumber
\end{eqnarray}
     Hence it is clear that the tortion $S_{\omega\nu}{^{\mu}} $
     is nonzero.

     Thus, the imbedding described in Sec.2 generates in
     $^kV_4 $ a geometry with tortion and a nonzero
     covariant derivative of the metric tensor.

     n8. The test particle motion equations
     are now considered. It is desirable for these
     equations to coincide with the geodesic equations
     in $^kV_4 $. Then these equations should take the form
\begin{equation}
\label{e9}
D p^{\mu}\: =\: - ^{(k)}\Gamma^{\mu}_{\nu o} p^o dx^{\nu}\: ,
 \; \; (p^{\mu}\: =\: m dx^{\mu}/d\tau)\: .
\end{equation}
     Taking this into account, the condition
     $dx_i dp^i = dx^o dp^o $ leads to the equation
\begin{equation}
\label{e10}
 {^{(k)}}S^j_{\nu o} dx^{\nu} dx_j\: =\:
 \Big(\Big\{^o_{\nu o}\Big\} + {^{(k)}}S^o_{\nu o}\Big)
 dx^{\nu} dx^o - \Big\{^j_{\nu o}\Big\} dx^{\nu} dx_j\: ,
\end{equation}
     which, together with Eq.(\ref{e8}), describes all the
     nonzero components of ${^{(k)}}S^{\mu}_{\omega\nu} $.

     Further, it is readily evident that, if the components
     ${^{(k)}}S^o_{\nu o} $ are chosen in the form
\begin{equation}
\label{e11}
 {^{(k)}}S^o_{\nu o}\: =\:
 [\partial_{\nu} \ln ((1 - {^{(k)}}g_{o o})/{^{(k)}}g_{o o})]/2\: ,
\end{equation}
     the four-momentum $p^o $ component is found to be
\begin{equation}
\label{e12}
 p^o\: =\: C_1(1 - {^{(k)}}g_{o o})^{-1/2}\: ,
\end{equation}
     where $C_1 = const $. Assuming that $C_1 = m c $, it is found
     that $d\tau = (1 - {^{(k)}}g_{o o})^{-1/2} dt $.

     Hence, Eqs.(\ref{e10}) and (\ref{e11}) are the necessary and
     sufficient conditions for the motion equations (\ref{e9})
     to be noncontradictory.

     Thus, the classical particle trajectories in the
     potential fields specified with respect to a definite
     IF may be represented as geodesic lines lying on
     isotropic surfaces of some configurational space
     $^kV_4 $ the connection of which has tortion, while
     the transference is nonmetric. The geometry of the space
     $^kV_4 $ has the distinctive property that the magnitude of
     the nonmetricity of the transfer and the tortion are
     determined by specifying the metric coefficient
     ${^{(k)}}g_{o o} $ under the condition that the mixed
     components ${^{(k)}}g_{o i} \equiv 0 $.

\end{document}